\documentclass[a4paper,11pt]{article}

\makeatletter

\gdef\@fpheader{\newline}
\gdef\@journal{jhep}

\RequirePackage{amsmath}
\RequirePackage{amssymb}
\RequirePackage{epsfig}
\RequirePackage{CJK}
\RequirePackage{graphicx}
\RequirePackage[numbers,sort&compress]{natbib}
\RequirePackage{color}
\RequirePackage[dvipdfmx
,colorlinks=true
,urlcolor=blue
,anchorcolor=blue
,citecolor=blue
,filecolor=blue
,linkcolor=blue
,menucolor=blue
,pagecolor=blue
,linktocpage=true
,pdfproducer=medialab
,pdfa=true
,CJKbookmarks
]{hyperref}

\newif\ifnotoc\notocfalse
\newif\ifemailadd\emailaddfalse
\newif\iftoccontinuous\toccontinuousfalse

\def\@subheader{\@empty}
\def\@keywords{\@empty}
\def\@abstract{\@empty}
\def\@xtum{\@empty}
\def\@dedicated{\@empty}
\def\@arxivnumber{\@empty}
\def\@collaboration{\@empty}
\def\@collaborationImg{\@empty}
\def\@proceeding{\@empty}
\def\@preprint{\@empty}

\newcommand{\subheader}[1]{\gdef\@subheader{#1}}
\newcommand{\keywords}[1]{\if!\@keywords!\gdef\@keywords{#1}\else%
\PackageWarningNoLine{\jname}{Keywords already defined.\MessageBreak Ignoring last definition.}\fi}
\renewcommand{\abstract}[1]{\gdef\@abstract{#1}}
\newcommand{\dedicated}[1]{\gdef\@dedicated{#1}}
\newcommand{\arxivnumber}[1]{\gdef\@arxivnumber{#1}}
\newcommand{\proceeding}[1]{\gdef\@proceeding{#1}}
\newcommand{\xtumfont}[1]{\textsc{#1}}
\newcommand{\correctionref}[3]{\gdef\@xtum{\xtumfont{#1} \href{#2}{#3}}}
\newcommand\jname{JHEP}
\newcommand\acknowledgments{\section*{Acknowledgments}}

\newcommand\preprint[1]{\gdef\@preprint{\hfill #1}}

\newcommand\note[2][]{%
\if!#1!%
\stepcounter{footnote}\footnotetext{#2}%
\else%
{\renewcommand\thefootnote{#1}%
\footnotetext{#2}}%
\fi}

\newtoks\auth@toks
\renewcommand{\author}[2][]{%
  \if!#1!%
    \auth@toks=\expandafter{\the\auth@toks#2\ }%
  \else
    \auth@toks=\expandafter{\the\auth@toks#2$^{#1}$\ }%
  \fi
}

\newtoks\affil@toks\newif\ifaffil\affilfalse
\newcommand{\affiliation}[2][]{%
\affiltrue
  \if!#1!%
    \affil@toks=\expandafter{\the\affil@toks{\item[]#2}}%
  \else
    \affil@toks=\expandafter{\the\affil@toks{\item[$^{#1}$]#2}}%
  \fi
}

\newtoks\email@toks\newcounter{email@counter}%
\newcommand{\emailAdd}[1]{%
\emailaddtrue%
\ifnum\theemail@counter>0\email@toks=\expandafter{\the\email@toks, \@email{#1}}%
\else\email@toks=\expandafter{\the\email@toks\@email{#1}}%
\fi\stepcounter{email@counter}}
\newcommand{\@email}[1]{\href{mailto:#1}{\tt #1}}

\newcommand*\collaboration[1]{\gdef\@collaboration{#1}}
\newcommand*\collaborationImg[2][]{\gdef\@collaborationImg{#2}}

\newcommand\afterLogoSpace{\smallskip}
\newcommand\afterSubheaderSpace{\vskip3pt plus 2pt minus 1pt}
\newcommand\afterProceedingsSpace{\vskip21pt plus0.4fil minus15pt}
\newcommand\afterTitleSpace{\vskip23pt plus0.06fil minus13pt}
\newcommand\afterRuleSpace{\vskip23pt plus0.06fil minus13pt}
\newcommand\afterCollaborationSpace{\vskip3pt plus 2pt minus 1pt}
\newcommand\afterCollaborationImgSpace{\vskip3pt plus 2pt minus 1pt}
\newcommand\afterAuthorSpace{\vskip5pt plus4pt minus4pt}
\newcommand\afterAffiliationSpace{\vskip3pt plus3pt}
\newcommand\afterEmailSpace{\vskip16pt plus9pt minus10pt\filbreak}
\newcommand\afterXtumSpace{\par\bigskip}
\newcommand\afterAbstractSpace{\vskip16pt plus9pt minus13pt}
\newcommand\afterKeywordsSpace{\vskip16pt plus9pt minus13pt}
\newcommand\afterArxivSpace{\vskip3pt plus0.01fil minus10pt}
\newcommand\afterDedicatedSpace{\vskip0pt plus0.01fil}
\newcommand\afterTocSpace{\bigskip\medskip}
\newcommand\afterTocRuleSpace{\bigskip\bigskip}
\newlength{\affiliationsSep}
\newcommand\beforetochook{\pagestyle{myplain}\pagenumbering{roman}}

\DeclareFixedFont\trfont{OT1}{phv}{b}{sc}{11}

\renewcommand\maketitle{
\pagestyle{empty}
\thispagestyle{titlepage}
\setcounter{page}{0}
\noindent{\small\scshape\@fpheader}\@preprint\par
\afterLogoSpace
\if!\@subheader!\else\noindent{\trfont{\@subheader}}\fi
\afterSubheaderSpace
\if!\@proceeding!\else\noindent{\sc\@proceeding}\fi
\afterProceedingsSpace
{\LARGE\flushleft\sffamily\bfseries\@title\par}
\afterTitleSpace
\hrule height 1.5\p@%
\afterRuleSpace
\if!\@collaboration!\else
{\Large\bfseries\sffamily\raggedright\@collaboration}\par
\afterCollaborationSpace
\fi
\if!\@collaborationImg!\else
{\normalsize\bfseries\sffamily\raggedright\@collaborationImg}\par
\afterCollaborationImgSpace
\fi
{\bfseries\raggedright\sffamily\the\auth@toks\par}
\afterAuthorSpace
\ifaffil\begin{list}{}{%
\setlength{\leftmargin}{0.28cm}%
\setlength{\labelsep}{0pt}%
\setlength{\itemsep}{\affiliationsSep}%
\setlength{\topsep}{-\parskip}}
\itshape\small%
\the\affil@toks
\end{list}\fi
\afterAffiliationSpace
\ifemailadd %% if emailadd is true
\noindent\hspace{0.28cm}\begin{minipage}[l]{.9\textwidth}
\begin{flushleft}
\textit{E-mail:} \the\email@toks
\end{flushleft}
\end{minipage}
\else %% if emailaddfalse do nothing
\PackageWarningNoLine{\jname}{E-mails are missing.\MessageBreak Plese use \protect\emailAdd\space macro to provide e-mails.}
\fi
\afterEmailSpace
\if!\@xtum!\else\noindent{\@xtum}\afterXtumSpace\fi
\if!\@abstract!\else\noindent{\renewcommand\baselinestretch{.9}\textsc{Abstract:}}\ \@abstract\afterAbstractSpace\fi
\if!\@keywords!\else\noindent{\textsc{Keywords:}} \@keywords\afterKeywordsSpace\fi
\if!\@arxivnumber!\else\noindent{\textsc{ArXiv ePrint:}} \href{http://arxiv.org/abs/\@arxivnumber}{\@arxivnumber}\afterArxivSpace\fi
\if!\@dedicated!\else\vbox{\small\it\raggedleft\@dedicated}\afterDedicatedSpace\fi
\ifnotoc\else
\iftoccontinuous\else\newpage\fi
\beforetochook\hrule
\tableofcontents
\afterTocSpace
\hrule
\afterTocRuleSpace
\fi
\setcounter{footnote}{0}
\pagestyle{myplain}\pagenumbering{arabic}
} % close the \renewcommand\maketitle{

\renewcommand{\baselinestretch}{1.1}\normalsize
\divide\@tempdima\baselineskip
\@tempcnta=\@tempdima
\skip\@mpfootins = \skip\footins
\renewcommand{\@dotsep}{10000}

\newcommand\ps@myplain{
\pagenumbering{arabic}
\renewcommand\@oddfoot{\hfill-- \thepage\ --\hfill}
\renewcommand\@oddhead{}}
\let\ps@plain=\ps@myplain

\newcommand\ps@titlepage{\renewcommand\@oddfoot{}\renewcommand\@oddhead{}}

\numberwithin{equation}{section}

\renewcommand\section{\@startsection{section}{1}{\z@}%
                                   {-3.5ex \@plus -1.3ex \@minus -.7ex}%
                                   {2.3ex \@plus.4ex \@minus .4ex}%
                                   {\normalfont\large\bfseries}}
\renewcommand\subsection{\@startsection{subsection}{2}{\z@}%
                                   {-2.3ex\@plus -1ex \@minus -.5ex}%
                                   {1.2ex \@plus .3ex \@minus .3ex}%
                                   {\normalfont\normalsize\bfseries}}
\renewcommand\subsubsection{\@startsection{subsubsection}{3}{\z@}%
                                   {-2.3ex\@plus -1ex \@minus -.5ex}%
                                   {1ex \@plus .2ex \@minus .2ex}%
                                   {\normalfont\normalsize\bfseries}}
\renewcommand\paragraph{\@startsection{paragraph}{4}{\z@}%
                                   {1.75ex \@plus1ex \@minus.2ex}%
                                   {-1em}%
                                   {\normalfont\normalsize\bfseries}}
\renewcommand\subparagraph{\@startsection{subparagraph}{5}{\parindent}%
                                   {1.75ex \@plus1ex \@minus .2ex}%
                                   {-1em}%
                                   {\normalfont\normalsize\bfseries}}

\def\fnum@figure{\textbf{\figurename\nobreakspace\thefigure}}
\def\fnum@table{\textbf{\tablename\nobreakspace\thetable}}

\long\def\@makecaption#1#2{%
  \vskip\abovecaptionskip
  \sbox\@tempboxa{\small #1. #2}%
  \ifdim \wd\@tempboxa >\hsize
    \small #1. #2\par
  \else
    \global \@minipagefalse
    \hb@xt@\hsize{\hfil\box\@tempboxa\hfil}%
  \fi
  \vskip\belowcaptionskip}

\renewenvironment{thebibliography}[1]{%
\begin{oldthebibliography}{#1}%
\small%
\raggedright%
\setlength{\itemsep}{5pt plus 0.2ex minus 0.05ex}%
}%
{%
\end{oldthebibliography}%
}

\makeatother

\usepackage[T1]{fontenc} % if needed
\usepackage{romannum} 
\usepackage{CJK}

\begin{document} %正文开始
\begin{CJK*}{GBK}{song}

%%%%%%%%题目作者%%%%%%%%%%%%%%%%%%%%%%%%%%%%%%%%%%%%%%%%%%%%%%%%

\title{{\boldmath A duality of fields} \\  }

% more complex case: 4 authors, 3 institutions, 2 footnotes
\author[a,b]{Wen-Du Li}
%\author[a,b]{Yu-Zhu Chen,}
\author[b*]{and Wu-Sheng Dai}\note{daiwusheng@tju.edu.cn.}

% The "\note" macro will give a warning: "Ignoring empty anchor..."
% you can safely ignore it.

\affiliation[a]{Theoretical Physics Division, Chern Institute of Mathematics, Nankai University, Tianjin, 300071, P. R. China}
\affiliation[b]{Department of Physics, Tianjin University, Tianjin 300350, P.R. China}

%\affiliation[c]{LiuHui Center for Applied Mathematics, Nankai University \& Tianjin University, Tianjin 300072, P.R. China}
%\affiliation[c]{DP School}

% e-mail addresses: one for each author, in the same order as the authors
%\emailAdd{Ccc@one.edu.cn}
%\emailAdd{second@asas.edu}
%\emailAdd{daiwusheng@tju.edu.cn}
%\emailAdd{fourth@one.univ}

%\title{\boldmath A title with some math: $x=1$}
%% %simple case: 2 authors, same institution
%% \author{A. Uthor}
%% \author{and A. Nother Author}
%% \affiliation{Institution,\\Address, Country}

% more complex case: 4 authors, 3 institutions, 2 footnotes
%\author[a,b,1]{F. Irst,\note{Corresponding author.}}
%\author[c]{S. Econd,}
%\author[a,2]{T. Hird\note{Also at Some University.}}
%\author[a,2]{and Fourth}

% The "\note" macro will give a warning: "Ignoring empty anchor..."
% you can safely ignore it.

%\affiliation[a]{One University,\\some-street, Country}
%\affiliation[b]{Another University,\\different-address, Country}
%\affiliation[c]{A School for Advanced Studies,\\some-location, Country}

% e-mail addresses: one for each author, in the same order as the authors

%\emailAdd{first@one.univ}
%\emailAdd{second@asas.edu}
%\emailAdd{third@one.univ}
%\emailAdd{fourth@one.univ}

%\date{date}

%\end{CJK*}

%%%%%%%%%%%%%%%%%%%%%%%%%%%%%%%%%%%%%%%%%%%%%%%%%%%%%%%%%%%%%%%%

%%%%%%%%%%%%%摘要和关键字%%%%%%%%%%%%%%%%%%%%%%%%%%%%%%%%%%%%%%%

\abstract{It is shown that there exists a duality among fields. If a field is dual to
another field, the solution of the field can be obtained from the dual field
by the duality transformation. We give a general result on the dual fields.
Different fields may have different numbers of dual fields, e.g., the free
field and the $\phi^{4}$-field are self-dual, the $\phi^{n}$-field has one
dual field, a field with an $n$-term polynomial potential has $n+1$ dual
fields, and a field with a nonpolynomial potential may have infinite number of
dual fields. All fields which are dual to each other form a duality family.
This implies that the field can be classified in the sense of duality, or, the
duality family defines a duality class. Based on the duality relation, we can
construct a high-efficiency approach for seeking the solution of field
equations: solving one field in the duality family, all solutions of other
fields in the family are obtained immediately by the duality transformation.
As examples, we consider some $\phi^{n}$-fields, general polynomial-potential
fields, and the sine-Gordon field.
}

%\keywords{......}

%%%%%%%%%%%%%%%%%%%%%%%%%%%%%%%%%%%%%%%%%%%%%%%%%%%%%%%%%%%%%%%%

%%%%%%%%%%%%%%%正文%%%%%%%%%%%%%%%%%%%%%%%%%%%%%%%%%%%%%%%%%%%%%

\maketitle %生成题目

\flushbottom
%(正文开始) ――――――――――――――――――――――――――――――――――――――――――――――――――

\section{Introduction}

In this paper, we show that there exists a duality between fields.
Furthermore, we show that the duality can serve as a method of solving field equations.

Consider a scalar field with the potential $V\left(  \phi\right)  $. The
Lagrangian is%
\begin{equation}
\mathcal{L}=\frac{1}{2}\partial_{\mu}\phi\partial^{\mu}\phi-\frac{1}{2}%
m^{2}\phi^{2}-V\left(  \phi\right)
\end{equation}
and the field equation is%
\begin{equation}
\square\phi+m^{2}\phi+\frac{\partial V\left(  \phi\right)  }{\partial\phi}=0.
\label{fieldeq}%
\end{equation}
It will be shown that the field with the potential $V\left(  \phi\right)  $
may have dual fields determined by some other potentials.

A field may have different numbers of dual fields. A $\phi^{n}$-field has one
dual field, a field with an $n$-term polynomial potential has $n+1$ dual
fields, a field with a nonpolynomial potential may have infinite number of
dual fields, and there are also self-dual fields. Dual fields form a duality
family; all fields in the family are dual to each other. This allows us to
classify fields based on the duality: a duality family forms a duality class.

The duality can serve as a high-efficiency approach for seeking the solution
of field equations. The duality relation relates a field and its dual field,
so the duality relation can be used to find the solution of a field equation
from the solution of the field equation of its dual field. In a duality
family, if one field in the family is solved, then the solution of all other
fields in the family can be obtained immediately by the duality
transformation. Concretely, we construct the dual field of the $\phi^{n}%
$-field, the general-polynomial-field ($V\left(  \phi\right)  $ is a general
polynomial of $\phi$), and the sine-Gordon field.

In physics, the duality plays an important role. The duality bridges two
different physical systems and reveals inherent relations in physics. Various
dualities are found in many physical problems. The $S$-duality (strong--weak
duality) relates a strongly coupled theory to an equivalent theory with a
small coupling constant, such as the duality between two perturbed quantum
many-body systems \cite{zarei2017strong}, the $(2+1)$-dimensional duality of
free Dirac or Majorana fermions and strongly-interacting bosonic
Chern-Simons-matter theories \cite{chen2018strong}, the electric-magnetic
duality \cite{bae2001electric,seiberg1995electric,hatsuda1999electric}, the
duality of higher spin free massless gauge fields \cite{boulanger2003note},
the duality in the couple of gauge field to gravity
\cite{igarashi1998electric}, and in string theory
\cite{duff1994four,font1990strong}. The AdS/CFT duality\ has been found in
many physical areas
\cite{maldacena1997large,witten1998anti,witten1998anti2,aharony2000large,d2004supersymmetric}%
. The fluid/gravity duality bridges fluid systems which is described by the
Navier-Stokes equation and spacetime which is described by the Einstein
equation
\cite{bredberg2012navier,hubeny2011fluid,compere2011holographic,hao2015flat,quiros2000dual,bhattacharyya2008nonlinear,bhattacharyya2009forced,bhattacharyya2008conformal,ashok2014forced,bhattacharyya2009incompressible,compere2012relativistic,pinzani2015towards,wu2013fluid}%
. The gravoelectric duality is another duality of spacetime
\cite{dadhich2002most,nouri1999spacetime,dadhich2000electromagnetic,dadhich1998duality,dadhich2000gravoelectric}%
. In string theory there is an T-duality
\cite{alvarez1995introduction,giveon1994target}.

In section \ref{phin}, we consider the duality of the $\phi^{n}$-field. In
section \ref{polynomial}, we consider the duality of the general polynomial
potential. In section \ref{sine-Gordon}, we consider the duality of the
sine-Gordon field. In section \ref{general}, we discuss an approach of the
construction of dual fields. In section \cite{generalpotential}, we give a
general result on dual fields. In section \ref{example}, we provide some
examples. The conclusion is given in section \ref{conclusion}. Moreover, we
add an appendix on the solution of field equations.

\section{The $\phi^{n}$-field \label{phin}}

In this section, we consider the duality of the $\phi^{n}$-field.

\subsection{The duality}

\textit{Two scalar fields }$\phi\left(  x^{\mu}\right)  $ and $\varphi\left(
y^{\mu}\right)  $\textit{\ with the potentials}%
\begin{align}
V\left(  \phi\right)   &  =\lambda\phi^{a},\label{Vphi}\\
U\left(  \varphi\right)   &  =\eta\varphi^{A} \label{Uphiw}%
\end{align}
\textit{are dual to each other if}%
\begin{equation}
\frac{2}{2-a}=\frac{2-A}{2}. \label{power}%
\end{equation}
\textit{The dual fields are related by the following duality relations:}%
\begin{align}
\phi &  \rightarrow\varphi^{\frac{a}{2-a}},\label{phitans}\\
x^{\mu}  &  \rightarrow\frac{2}{2-a}y^{\mu},\text{ \ }\mu=0,1,\ldots
\label{xtans}%
\end{align}
\textit{and}%
\begin{align}
\lambda &  \rightarrow-\mathcal{G},\label{xie}\\
G  &  \rightarrow-\eta, \label{Eeta}%
\end{align}
\textit{where }%
\begin{align}
G  &  =\frac{1}{2}\partial_{\mu}\phi\partial^{\mu}\phi+\frac{1}{2}m^{2}%
\phi^{2}+\lambda\phi^{a},\label{Ephi}\\
\mathcal{G}  &  =\frac{1}{2}\partial_{\mu}\varphi\partial^{\mu}\varphi
+\frac{1}{2}m^{2}\varphi^{2}+\eta\varphi^{A} \label{Evarphi}%
\end{align}
\textit{ are two Lorentz scalars corresponding to }$\phi$\textit{ and
}$\varphi$\textit{, respectively.}

The duality relation can be verified directly.

The field equation of the potential (\ref{Vphi}) is
\begin{equation}
\square\phi+m^{2}\phi+a\lambda\phi^{a-1}=0. \label{phieqa}%
\end{equation}
The duality transformations (\ref{phitans}) and (\ref{xtans}) give%
\begin{equation}
\square\phi\rightarrow\frac{2-a}{2}\varphi^{\frac{a}{2-a}}\square\varphi
+\frac{a}{2}\varphi^{\frac{2}{2-a}-2}\partial_{\mu}\varphi\partial^{\mu
}\varphi. \label{2dtansx}%
\end{equation}
Substituting the transformation (\ref{2dtansx}) into the field equation
(\ref{phieqa}) gives%
\begin{equation}
\frac{2-a}{2}\varphi^{\frac{a}{2-a}}\square\varphi+\frac{a}{2}\varphi
^{\frac{2}{2-a}-2}\partial_{\mu}\varphi\partial^{\mu}\phi+m^{2}\varphi
^{\frac{2}{2-a}}+a\lambda\varphi^{\frac{2\left(  a-1\right)  }{2-a}}=0.
\label{equatans1}%
\end{equation}
The coupling constant $\lambda$ in the potential $V\left(  \phi\right)  $, by
the duality transformation (\ref{xie}), is replaced by:%
\begin{equation}
\lambda\rightarrow-\left(  \frac{1}{2}\partial_{\mu}\varphi\partial^{\mu
}\varphi+\frac{1}{2}m^{2}\varphi^{2}+\eta\varphi^{A}\right)  . \label{lbdG}%
\end{equation}
Using the duality relation (\ref{power}), we arrive at the field equation of
$\varphi$,%
\begin{equation}
\square\varphi+m^{2}\varphi+A\eta\varphi^{A-1}=0. \label{varphieqa}%
\end{equation}

\subsection{Solving field equations by means of the duality}

The duality relation bridges various fields and can serve as a method of
solving field equations. Once a field equation is solved, its dual field is
also solved.

It can be checked that the field equation with the potential (\ref{phieqa})
has an implicit solution (Appendix \ref{appendix}):%
\begin{equation}
\beta_{\mu}x^{\mu}+\int\frac{\sqrt{-\beta^{2}}}{\sqrt{2\left(  \frac{1}%
{2}m^{2}\phi^{2}+\lambda\phi^{a}-G\right)  }}d\phi=0, \label{solphia}%
\end{equation}
where $\beta_{\mu}$ is a constant.

Substituting the duality transformations (\ref{phitans}) and (\ref{xtans})
into the solution (\ref{solphia}) gives%
\begin{equation}
\beta_{\mu}y^{\mu}+\int\frac{\sqrt{-\beta^{2}}}{\sqrt{2\left[  \frac{1}%
{2}m^{2}\varphi^{2}+\left(  -G\varphi^{-\frac{2a}{2-a}}\right)  -\left(
-\lambda\right)  \right]  }}d\varphi=0. \label{solivarphiA1}%
\end{equation}
This is just a solution of the field equation with $U\left(  \varphi\right)
=-G\varphi^{-\frac{2a}{2-a}}$. By the duality relations (\ref{power}),
(\ref{xie}), (\ref{Eeta}), and Eq. (\ref{Evarphi}), we can see that this is
the solution of the field equation (\ref{varphieqa}):%
\begin{equation}
\beta_{\mu}y^{\mu}+\int\frac{\sqrt{-\beta^{2}}}{\sqrt{2\left(  \frac{1}%
{2}m^{2}\varphi^{2}+\eta\varphi^{A}-\mathcal{G}\right)  }}d\varphi=0.
\end{equation}

\section{The general polynomial potential \label{polynomial}}

In this section, we consider the duality between the field with general
polynomial potentials. The general polynomial is a superposition power series
containing arbitrary real-number powers.

\subsection{The duality}

\textit{Two scalar fields }$\phi\left(  x^{\mu}\right)  $ and $\varphi\left(
y^{\mu}\right)  $\textit{\ of the potentials}%
\begin{align}
V\left(  \phi\right)   &  =\lambda\phi^{a}+\sum_{n}\kappa_{n}\phi^{b_{n}%
},\label{GPPV}\\
U\left(  \varphi\right)   &  =\eta\varphi^{A}+\sum_{n}\sigma_{n}\varphi
^{B_{n}} \label{GPPU}%
\end{align}
\textit{are dual to each other if}%
\begin{align}
\frac{2-A}{2}  &  =\frac{2}{2-a},\label{aandA}\\
2-b_{n}  &  =\frac{2}{2-A}\left(  2-B_{n}\right)  \text{ or }\frac{2}%
{2-a}\left(  2-b_{n}\right)  =2-B_{n}. \label{bandB}%
\end{align}
\textit{The dual fields are related by the following duality relations:}%
\begin{align}
\phi &  \rightarrow\varphi^{\frac{2}{2-a}},\label{phitanspol}\\
x^{\mu}  &  \rightarrow\frac{2}{2-a}y^{\mu},\text{ \ }\mu=0,1,\ldots
\label{xtanspol}%
\end{align}
\textit{and}%
\begin{align}
\lambda &  \rightarrow-\mathcal{G},\label{lamG}\\
G  &  \rightarrow-\eta,\label{Geta}\\
\kappa_{n}  &  \rightarrow\sigma_{n}, \label{kapsig}%
\end{align}
\textit{where}%
\begin{align}
G  &  =\frac{1}{2}\partial_{\mu}\phi\partial^{\mu}\phi+\frac{1}{2}m^{2}%
\phi^{2}+V\left(  \phi\right)  ,\label{Gphin}\\
\mathcal{G}  &  =\frac{1}{2}\partial_{\mu}\varphi\partial^{\mu}\varphi
+\frac{1}{2}m^{2}\varphi^{2}+U\left(  \varphi\right)  \label{Ghua}%
\end{align}
\textit{ are two Lorentz scalars corresponding to }$\phi$\textit{ and
}$\varphi$\textit{, respectively.}

The field equation with the potential (\ref{GPPV}) is%
\begin{equation}
\square\phi+m^{2}\phi+a\lambda\phi^{a-1}+\sum_{n}b_{n}\kappa_{n}\phi^{b_{n}%
-1}=0. \label{phieqan}%
\end{equation}

The duality transformations (\ref{phitanspol}) and (\ref{xtanspol}) give%
\begin{equation}
\square\phi\rightarrow\frac{2-a}{2}\varphi^{\frac{a}{2-a}}\square\varphi
+\frac{a}{2}\varphi^{\frac{2}{2-a}-2}\partial_{\mu}\varphi\partial^{\mu
}\varphi. \label{2dtansxn}%
\end{equation}
Substituting the transformation into the field equation (\ref{phieqan}) gives
an equation of $\varphi$:%
\begin{equation}
\frac{2-a}{2}\varphi^{\frac{a}{2-a}}\square\varphi+\frac{a}{2}\varphi
^{\frac{2}{2-a}-2}\partial_{\mu}\varphi\partial^{\mu}\phi+m^{2}\varphi
^{\frac{2}{2-a}}+a\lambda\varphi^{\frac{2\left(  a-1\right)  }{2-a}}+\sum
_{n}b_{n}\kappa_{n}\varphi^{\frac{2\left(  b_{n}-1\right)  }{2-a}}=0.
\label{equatansn1}%
\end{equation}
The transformation of the coupling constant is given by the duality
transformation (\ref{lamG}):%
\begin{equation}
\lambda\rightarrow-\left(  \frac{1}{2}\partial_{\mu}\varphi\partial^{\mu
}\varphi+\frac{1}{2}m^{2}\varphi^{2}+\eta\varphi^{A}+\sum_{n}\sigma_{n}%
\varphi^{B_{n}}\right)  .
\end{equation}
Then we arrive at the field equation of $U\left(  \varphi\right)  =\eta
\varphi^{A}+\sum_{n}\sigma_{n}\phi^{B_{n}}$ with the duality relations
(\ref{aandA}), (\ref{bandB}), and (\ref{kapsig}):
\begin{equation}
\square\varphi+m^{2}\varphi+A\eta\varphi^{A-1}+\sum_{n}B_{n}\sigma_{n}%
\varphi^{B_{n}-1}=0.
\end{equation}

It can be seen from the potential (\ref{GPPV}) that in a general polynomial
potential $V\left(  \phi\right)  =\lambda\phi^{a}+\sum_{n}\kappa_{n}%
\phi^{b_{n}}$, every term in the potential can be chosen as the first term in
$V\left(  \phi\right)  $ to play the role of $\lambda\phi^{a}$. Different
choices give different dual fields. For an $n$-term polynomial potential,
there are $n$ choices. Therefore, all potentials who are dual to each other
form a duality family with $n+1$ members.

\subsection{Solving field equations by means of the duality}

Similarly, the field equation (\ref{phieqan}) has an implicit solution:%
\begin{equation}
\beta_{\mu}x^{\mu}+\int\frac{\sqrt{-\beta^{2}}}{\sqrt{2\left[  \frac{1}%
{2}m^{2}\phi^{2}+\left(  \lambda\phi^{a}+\sum_{n}\kappa_{n}\phi^{b_{n}%
}\right)  -G\right]  }}d\phi=0. \label{solpol}%
\end{equation}

Substituting the duality transformations (\ref{phitanspol}) and
(\ref{xtanspol}) into the solution (\ref{solpol}) and using the duality
relations (\ref{aandA}) and (\ref{bandB}), we arrive at%
\begin{equation}
\beta_{\mu}y^{\mu}+\int\frac{\sqrt{-\beta^{2}}}{\sqrt{2\left[  \frac{1}%
{2}m^{2}\varphi^{2}+\left(  -G\varphi^{-\frac{2a}{2-a}}+\sum_{n}\kappa
_{n}\varphi^{\frac{2b_{n}-2a}{2-a}}\right)  -\left(  -\lambda\right)  \right]
}}d\varphi=0. \label{solpol2}%
\end{equation}
This is just a solution of the field equation with $U\left(  \varphi\right)
=-G\varphi^{-\frac{2a}{2-a}}+\sum_{n}\kappa_{n}\varphi^{\frac{2b_{n}-2a}{2-a}%
}$. By the duality relations (\ref{aandA}), (\ref{bandB}), (\ref{lamG}),
(\ref{Geta}), (\ref{kapsig}), and Eq. (\ref{Ghua}), we can see that this is
the solution of the field equation with the potential (\ref{varphieqa}):%
\begin{equation}
\beta_{\mu}y^{\mu}+\int\frac{\sqrt{-\beta^{2}}}{\sqrt{2\left[  \frac{1}%
{2}m^{2}\varphi^{2}+U\left(  \varphi\right)  -\mathcal{G}\right]  }}%
d\varphi=0.
\end{equation}

Obviously, for an $n$-term polynomial potential, once a potential in the
duality family is solved, the other $n$ potentials are immediately solved.

\section{The sine-Gordon equation \label{sine-Gordon}}

The sine-Gordon potential is not a polynomial potential. We first expand the
sine-Gordon potential as a power series. Each term in the expansion is a power
function, which has a duality discussed in section \ref{phin}. Calculating the
duality of each term of the series and then summing up the series give the
dual potential of the sine-Gordon potential.

\subsection{The duality}

The sine-Gordon equation is \cite{Rajaraman1987solitons}%
\begin{equation}
\square\phi+\frac{m^{3}}{\sqrt{\lambda}}\sin\left(  \frac{\sqrt{\lambda}}%
{m}\phi\right)  =0. \label{SinGeq}%
\end{equation}
First rewrite the sine-Gordon Lagrangian, $\mathcal{L}=\frac{1}{2}%
\partial_{\mu}\phi\partial^{\mu}\phi-\frac{m^{4}}{\lambda}\left[
1-\cos\left(  \frac{\sqrt{\lambda}}{m}\phi\right)  \right]  $, as%
\begin{equation}
\mathcal{L}=\frac{1}{2}\partial_{\mu}\phi\partial^{\mu}\phi-\frac{1}{2}%
m^{2}\phi^{2}-V\left(  \phi\right)
\end{equation}
with
\begin{equation}
V\left(  \phi\right)  =\frac{m^{4}}{\lambda}\left[  1-\cos\left(  \frac
{\sqrt{\lambda}}{m}\phi\right)  \right]  -\frac{1}{2}m^{2}\phi^{2}.
\label{Vphimphi2}%
\end{equation}
Expanding $V\left(  \phi\right)  $ gives%
\begin{equation}
V\left(  \phi\right)  =-\frac{\lambda}{24}\phi^{4}-\sum_{n=3}^{\infty}%
\frac{\left(  -1\right)  ^{n}}{\left(  2n\right)  !}\frac{\lambda^{n-1}%
}{m^{2\left(  n-2\right)  }}\phi^{2n}. \label{Veff}%
\end{equation}
The leading term in Eq. (\ref{Veff}) is the $\phi^{4}$ term.

First regarding the expansion (\ref{Veff}) as a polynomial (though it is
indeed a series), we can obtain the dual potential by the duality
transformation of the general polynomial potential, Eqs. (\ref{aandA}),
(\ref{bandB}), (\ref{lamG}), (\ref{Geta}), and (\ref{kapsig}):%
\begin{align}
A  &  =4,\text{ \ }B_{n}=4-2n,\text{ \ }n=3,4,5,\ldots,\nonumber\\
\eta &  =-G,\text{ \ }\sigma_{n}=-\frac{\left(  -1\right)  ^{n}}{\left(
2n\right)  !}\frac{\lambda^{n-1}}{m^{2\left(  n-2\right)  }}. \label{sineGT}%
\end{align}
The duality transformations here, by Eqs. (\ref{phitanspol}) and
(\ref{xtanspol}), are
\begin{align}
\phi &  \rightarrow\varphi^{-1},\label{SinKphiphi}\\
x^{\mu}  &  \rightarrow-y^{\mu}. \label{SinKxy}%
\end{align}

The potential of the dual field can be obtained by substituting the duality
relation (\ref{sineGT}) into Eq. (\ref{GPPU}) and then summing up the series:%
\begin{align}
U\left(  \varphi\right)   &  =-G\varphi^{4}-\sum_{n=3}^{\infty}\frac{\left(
-1\right)  ^{n}}{\left(  2n\right)  !}\frac{\lambda^{n-1}}{m^{2\left(
n-2\right)  }}\varphi^{4-2n}\nonumber\\
&  =-G\varphi^{4}+\frac{m^{4}}{\lambda}\varphi^{4}\left(  1+\cos\frac
{\sqrt{\lambda}}{m\varphi}\right)  -\frac{1}{2}m^{2}\varphi^{2}.
\label{SinGUphi}%
\end{align}
The field equation of the dual field then reads%
\begin{equation}
\square\phi-4G\varphi^{3}+\frac{4m^{4}}{\lambda}\varphi^{3}\left(  1-\cos
\frac{\sqrt{\lambda}}{m\varphi}\right)  -\frac{m^{3}}{\sqrt{\lambda}}%
\varphi^{3}\sin\frac{\sqrt{\lambda}}{m\varphi}=0.
\end{equation}
Different solution of $\phi$ gives different $G$ and different $G$ gives
different dual potentials.

The expansion (\ref{Veff}) has infinite terms, so the duality family has
infinite members.

\subsubsection{The $1+n$-dimensional solution}

The sine-Gordon equation (\ref{SinGeq}) has a\ $1+n$-dimensional solution
\cite{Rajaraman1987solitons}
\begin{equation}
\phi=\frac{4m}{\sqrt{\lambda}}\operatorname{arccot}\left(  i\cot\left(
\frac{\beta_{\mu}x^{\mu}}{2\sqrt{-\beta^{2}}}m\right)  \right)  .
\label{sgslo}%
\end{equation}

Substituting the solution (\ref{sgslo}) into Eq. (\ref{Gphin}) gives
\begin{equation}
G=\frac{2m^{4}}{\lambda}.
\end{equation}

Then the potential of the dual field, by Eq. (\ref{SinGUphi}), reads
\begin{equation}
U\left(  \varphi\right)  =-\frac{m^{4}}{\lambda}\varphi^{4}\left(  1+\cos
\frac{\sqrt{\lambda}}{m\varphi}\right)  -\frac{1}{2}m^{2}\varphi^{2}.
\label{Uvarphi}%
\end{equation}
The field equation of the dual field is then%
\begin{equation}
\square\varphi+\frac{4m^{4}}{\lambda}\varphi^{3}\left(  1+\cos\frac
{\sqrt{\lambda}}{m\varphi}\right)  +\frac{m^{3}}{\sqrt{\lambda}}\varphi
^{2}\sin\frac{\sqrt{\lambda}}{m\varphi}=0.
\end{equation}

The solution of the dual field can be obtained by substituting the duality
transformations (\ref{phitanspol}) and (\ref{xtanspol}) into the solution
(\ref{sgslo}):%
\begin{equation}
\varphi=\left[  \frac{4m}{\sqrt{\lambda}}\operatorname{arccot}\left(
i\cot\left(  \frac{-\beta_{\mu}y^{\mu}}{2\sqrt{-\beta^{2}}}m\right)  \right)
\right]  ^{-1}.
\end{equation}

\subsubsection{The $1+1$-dimensional solution}

The sine-Gordon equation (\ref{SinGeq}) also has a $1+1$-dimensional solution
\cite{ablowitz1973method, ablowitz1991solitons},%
\begin{equation}
\phi=\frac{4m}{\sqrt{\lambda}}\arctan\left(  \frac{\cos\left(  \frac{\sqrt{2}%
}{2}mt\right)  }{\cosh\left(  \frac{\sqrt{2}}{2}mx\right)  }\right)  .
\label{sgslo1}%
\end{equation}

Substituting the solution (\ref{sgslo}) into (\ref{Gphin}) gives%
\begin{equation}
G=\frac{8m^{4}}{\lambda}\left[  \cos\left(  \sqrt{2}mt\right)  +\cosh\left(
\sqrt{2}mx\right)  +2\right]  ^{-1}.
\end{equation}

The potential of the dual field, by Eq. (\ref{SinGUphi}), reads%
\begin{equation}
U\left(  \varphi\right)  =-\frac{8m^{4}}{\lambda}\left[  \cos\left(  \sqrt
{2}mt\right)  +\cosh\left(  \sqrt{2}mx\right)  +2\right]  ^{-1}\varphi
^{4}+\frac{m^{4}}{\lambda}\varphi^{4}\left(  1-\cos\left(  \frac{\sqrt
{\lambda}}{m\varphi}\right)  \right)  -\frac{1}{2}m^{2}\varphi^{2}.
\label{Uvarphibian}%
\end{equation}
The field equation of the dual field is%
\begin{equation}
\square\varphi-\frac{32m^{4}}{\lambda}\left[  \cos\left(  \sqrt{2}mt\right)
+\cosh\left(  \sqrt{2}mx\right)  +2\right]  ^{-1}\varphi^{3}+\frac{4m^{4}%
}{\lambda}\varphi^{3}\left(  1-\cos\frac{\sqrt{\lambda}}{m\varphi}\right)
-\frac{m^{3}}{\sqrt{\lambda}}\varphi^{3}\sin\frac{\sqrt{\lambda}}{m\varphi}=0.
\label{Uvarphibianeq}%
\end{equation}

The solution of the dual field can be obtained by substituting the duality
transformations (\ref{SinKphiphi}) and (\ref{SinKxy}) into the solution
(\ref{sgslo1}):%
\begin{equation}
\varphi=\left[  \frac{4m}{\sqrt{\lambda}}\arctan\left(  \frac{\cos\left(
\frac{\sqrt{2}}{2}mt\right)  }{\cosh\left(  \frac{\sqrt{2}}{2}mx\right)
}\right)  \right]  ^{-1}.
\end{equation}
This is the solution of the field equation (\ref{Uvarphibianeq}).

\subsection{The duality family of the sine-Gordon field}

In the above, by regarding the expansion of the sine-Gordon potential as a
\textquotedblleft polynomial potential\textquotedblright, we obtain the dual
potential of the sine-Gordon potential. In section \ref{polynomial}, we write
a polynomial potential in the form of $V\left(  \phi\right)  =\lambda\phi
^{a}+\sum_{n}\kappa_{n}\phi^{b_{n}}$, in which the first term $\lambda\phi
^{a}$ is arbitrarily chosen from the polynomial and the other terms in the
polynomial are incorporated into the sum. However, every term in $V\left(
\phi\right)  $ can be chosen as the first term and different choices lead to
different dual potentials. In the following, we give a general discussion on
the dual potential of the sine-Gordon potential by choosing different terms in
the expansion as the first term.

Rewriting the expansion of $V\left(  \phi\right)  $, Eq. (\ref{Vphimphi2}),
as
\begin{equation}
V\left(  \phi\right)  =-\frac{\left(  -1\right)  ^{\chi}}{\left(
2\chi\right)  !}\frac{\lambda^{\chi-1}}{m^{2\left(  \chi-2\right)  }}%
\phi^{2\chi}-\sum_{n=2\left(  n\neq\chi\right)  }^{\infty}\frac{\left(
-1\right)  ^{n}}{\left(  2n\right)  !}\frac{\lambda^{n-1}}{m^{2\left(
n-2\right)  }}\phi^{2n},\ \chi=2,3,4\cdots. \label{Veff1}%
\end{equation}
It can seen that the expansion (\ref{Vphimphi2}) is just the special case of
Eq. (\ref{Veff1}) with $\chi=2$.

Similarly, regarding the expansion (\ref{Veff1}) as a "polynomial", we obtain
the dual potential by the duality transformation of the general polynomial
potential, Eqs. (\ref{aandA}), (\ref{bandB}), (\ref{lamG}), (\ref{Geta}), and
(\ref{kapsig}):%
\begin{align}
A  &  =\frac{2\chi}{\chi-1},\text{ \ }B_{n}=\frac{2\left(  n-\chi\right)
}{1-\chi},\text{ \ }\nonumber\\
\eta &  =-G,\text{ \ }\sigma_{n}=-\frac{\left(  -1\right)  ^{n}}{\left(
2n\right)  !}\frac{\lambda^{n-1}}{m^{2\left(  n-2\right)  }},\text{\ }%
n=2,3,4,\ldots,\text{ and }n\neq\chi. \label{sigkappa}%
\end{align}
The duality transformations (\ref{phitanspol}) and (\ref{xtanspol}) then
become%
\begin{align}
\phi &  \rightarrow\varphi^{\frac{1}{1-\chi}},\label{phiphi}\\
x^{\mu}  &  \rightarrow\frac{1}{1-\chi}y^{\mu}. \label{xtoy}%
\end{align}

Substituting the duality relation into Eq. (\ref{GPPU}) and summing up the
series give the dual potential of $V\left(  \phi\right)  $%

\begin{align}
U\left(  \varphi\right)   &  =-G\varphi^{\frac{2\chi}{\chi-1}}-\sum
_{n=2^{\prime}}^{\infty}\frac{\left(  -1\right)  ^{n}}{\left(  2n\right)
!}\frac{\lambda^{n-1}}{m^{2\left(  n-2\right)  }}\varphi^{\frac{2\left(
n-\chi\right)  }{1-\chi}}\\
&  =-G\varphi^{\frac{2\chi}{\chi-1}}-\frac{m^{4}}{\lambda}\varphi^{\frac
{2\chi}{\chi-1}}\left[  \cos\left(  \frac{\sqrt{\lambda}}{m}\varphi^{\frac
{1}{1-\chi}}\right)  -1\right]  -\frac{1}{2}m^{2}\varphi^{2}.
\end{align}
The field equation is then%
\begin{equation}
\square\varphi-\frac{2\chi}{\chi-1}G\varphi^{\frac{\chi+1}{\chi-1}}%
-\frac{2\chi}{\chi-1}\frac{m^{4}}{\lambda}\varphi^{\frac{\chi+1}{\chi-1}%
}\left[  \cos\left(  \frac{\sqrt{\lambda}}{m}\varphi^{\frac{1}{1-\chi}%
}\right)  -1\right]  +\frac{1}{1-\chi}\frac{m^{3}}{\sqrt{\lambda}}%
\varphi^{\frac{\chi}{\chi-1}}\sin\left(  \frac{\sqrt{\lambda}}{m}%
\varphi^{\frac{1}{1-\chi}}\right)  =0.
\end{equation}

Substituting the solution (\ref{sgslo}) into Eq. (\ref{Gphin}) gives%
\begin{equation}
G=\frac{2m^{4}}{\lambda}.
\end{equation}
Then we arrive at%
\begin{align}
U\left(  \varphi\right)   &  =-\frac{2m^{4}}{\lambda}\varphi^{\frac{2\chi
}{\chi-1}}-\sum_{n=2^{\prime}}^{\infty}\frac{\left(  -1\right)  ^{n}}{\left(
2n\right)  !}\frac{\lambda^{n-1}}{m^{2\left(  n-2\right)  }}\varphi
^{\frac{2\left(  n-\chi\right)  }{1-\chi}}\nonumber\\
&  =-\frac{m^{4}}{\lambda}\varphi^{\frac{2\chi}{\chi-1}}\left(  1+\cos
\frac{\sqrt{\lambda}}{m}\varphi^{\frac{1}{1-\chi}}\right)  -\frac{1}{2}%
m^{2}\varphi^{2}-\left(  -\frac{\lambda}{m^{4}}\right)  ^{\chi-1}%
\frac{m^{2\chi}}{\left(  2\chi\right)  !}.
\end{align}
The field equation of the dual field then reads
\[
\square\varphi-\frac{2\chi}{\chi-1}\frac{m^{4}}{\lambda}\varphi^{\frac{\chi
+1}{\chi-1}}\left[  \cos\left(  \frac{\sqrt{\lambda}}{m}\varphi^{\frac
{1}{1-\chi}}\right)  +1\right]  +\frac{1}{1-\chi}\frac{m^{3}}{\sqrt{\lambda}%
}\varphi^{\frac{\chi}{\chi-1}}\sin\left(  \frac{\sqrt{\lambda}}{m}%
\varphi^{\frac{1}{1-\chi}}\right)  =0.
\]

Substituting the duality relations (\ref{phiphi}) and (\ref{xtoy}) into the
solution (\ref{sgslo}) gives the solution of the dual potential:%
\begin{equation}
\varphi=\left[  \frac{4m}{\sqrt{\lambda}}\operatorname{arccot}\left(
i\cot\left(  -\frac{1}{1-\chi}\frac{\beta_{\mu}y^{\mu}}{2\sqrt{-\beta^{2}}%
}m\right)  \right)  \right]  ^{1-\chi}.
\end{equation}

\section{Fields with general potentials \label{generalpotential}}

For nonpolynomial potentials, like that in the case of the sine-Gordon fields,
we can first expand the potential as a series, then construct the dual field
with the help of the duality relation of the general polynomial potential
given in section \ref{polynomial}, and sum up the series.

Expand the potential $V\left(  \phi\right)  $ as%
\begin{equation}
V\left(  \phi\right)  =\lambda\phi^{a}+\sum_{n}\kappa_{n}\phi^{b_{n}}.
\label{Vphiseries}%
\end{equation}
Using the duality relations (\ref{aandA}), (\ref{bandB}), (\ref{lamG}),
(\ref{Geta}), and (\ref{kapsig}), we obtain the series of the dual potential,%
\begin{align}
U\left(  \varphi\right)   &  =-G\varphi^{\frac{2a}{a-2}}+\sum_{n}\kappa
_{n}\varphi^{\frac{2a}{a-2}-\frac{2b_{n}}{a-2}}\nonumber\\
&  =-G\left(  \varphi^{\frac{2}{2-a}}\right)  ^{-a}+\left(  \varphi^{\frac
{2}{2-a}}\right)  ^{-a}\sum_{n}\kappa_{n}\left(  \varphi^{\frac{2}{2-a}%
}\right)  ^{b_{n}}. \label{Uvarphiseries1}%
\end{align}
By Eq. (\ref{Vphiseries}) we have $\sum_{n}\kappa_{n}\phi^{b_{n}}=V\left(
\phi\right)  -\lambda\phi^{a}$, so the series in Eq. (\ref{Uvarphiseries1})
can be summed up as
\begin{align}
U\left(  \varphi\right)   &  =-G\left(  \varphi^{\frac{2}{2-a}}\right)
^{-a}+\left(  \varphi^{\frac{2}{2-a}}\right)  ^{-a}\left[  V\left(
\varphi^{\frac{2}{2-a}}\right)  -\lambda\left(  \varphi^{\frac{2}{2-a}%
}\right)  ^{a}\right] \nonumber\\
&  =-G\varphi^{\frac{2a}{a-2}}+\varphi^{\frac{2a}{a-2}}V\left(  \varphi
^{\frac{2}{2-a}}\right)  -\lambda.
\end{align}
This is a general result on dual fields.

\section{ Constructing the dual fields from the solution \label{general}}

In this section, we discuss an approach of constructing the dual field from
the solution.

In the implicit solution of the field equation (\ref{fieldeq}),%
\begin{equation}
\beta_{\mu}x^{\mu}+\int\frac{\sqrt{-\beta^{2}}}{\sqrt{2\left[  \frac{1}%
{2}m^{2}\phi^{2}+V\left(  \phi\right)  -G\right]  }}d\phi=0, \label{solution}%
\end{equation}
by writing the duality transformation in the following general form,
\begin{align}
\phi &  \rightarrow f\left(  \varphi\right)  ,\\
x^{\mu}  &  \rightarrow g^{\mu}\left(  y\right)  ,
\end{align}
the solution (\ref{solution}) becomes%
\begin{equation}
\beta_{\mu}g^{\mu}\left(  y\right)  +\int\frac{\sqrt{-\beta^{2}}}%
{\sqrt{2\left[  \frac{1}{2}m^{2}f^{2}\left(  \varphi\right)  +V\left(
f\left(  \varphi\right)  \right)  -G\right]  }}f^{\prime}\left(
\varphi\right)  d\varphi=0. \label{dualtr1}%
\end{equation}

In order that Eq. (\ref{dualtr1}) is still a solution of a field equation, a
simple choice is%
\begin{equation}
g^{\mu}\left(  y\right)  =\chi y^{\mu} \label{gyxy}%
\end{equation}
with $\chi$ a constant. Then Eq. (\ref{dualtr1}) with the duality
transformation (\ref{gyxy}) becomes%
\begin{equation}
\beta_{\mu}y^{\mu}+\int\frac{\sqrt{-\beta^{2}}}{\sqrt{2\left\{  \frac{1}%
{2}m^{2}\chi2\frac{f^{2}\left(  \varphi\right)  }{\left[  f^{\prime}\left(
\varphi\right)  \right]  ^{2}}+\frac{\chi^{2}}{\left[  f^{\prime}\left(
\varphi\right)  \right]  ^{2}}V\left[  f\left(  \varphi\right)  \right]
-G\frac{\chi^{2}}{\left[  f^{\prime}\left(  \varphi\right)  \right]  ^{2}%
}\right\}  }}d\varphi=0. \label{dualtr2}%
\end{equation}
Requiring that Eq. (\ref{dualtr2}) is still a solution of the field equation
and comparing Eqs. (\ref{solphia}) and (\ref{dualtr2}), we can choose (1)
one\ of the three terms as the mass term which should be in proportion to
$\varphi^{2}$, (2) one of the other terms as the constant term, and (3) the
remaining term as the potential term $U\left(  \varphi\right)  $. Different
choices give different dual fields.

\section{The duality of polynomial fields: examples \label{example}}

In this section, we give some examples of the method of solving the field
equation by the duality transformation.

\subsection{The self-duality: free fields and $\phi^{4}$-fields}

The self-duality means that the dual field of a field is the field itself. The
free field and the $\phi^{4}$-field are self-dual. This can be seen directly
from the duality relation (\ref{power}): if $a=0$, then $A=0$; if $a=$ $4$,
then $A=4$.

Take the $\phi^{4}$-field as an example. The field equation of the $\phi^{4}%
$-field is
\begin{equation}
\square\phi+m^{2}\phi+4\lambda\phi^{3}=0, \label{p4}%
\end{equation}
which has a soliton solution \cite{polyanin2016handbook},
\begin{equation}
\phi=\frac{im}{2\sqrt{\lambda}}\tanh\left(  \alpha t+\beta x_{1}+\gamma
x_{2}-\frac{x_{3}}{2}\sqrt{4\alpha^{2}-4\beta^{2}-4\gamma^{2}-2m^{2}}%
+\delta\right)  . \label{p4sol}%
\end{equation}

For the $\phi^{4}$-field potential, by Eqs. (\ref{phitans}) and (\ref{xtans}),
the duality transformations are%
\begin{align}
\phi &  \rightarrow\varphi^{-1},\label{phi4phitrans}\\
x^{\mu}  &  \rightarrow-y^{\mu}. \label{xtoyphi4}%
\end{align}
The field equation (\ref{p4}) under the duality transformation becomes%
\begin{equation}
-\varphi^{-2}\square\varphi+2\varphi^{-3}\partial_{\mu}\varphi\partial^{\mu
}\phi+m^{2}\varphi^{-1}+4\lambda\varphi^{-3}=0. \label{P41}%
\end{equation}

The coupling constant $\lambda$, by the duality relation (\ref{xie}) and Eq.
(\ref{Evarphi}), should be replaced by%
\begin{equation}
\lambda\rightarrow-\frac{1}{2}\partial_{\mu}\varphi\partial^{\mu}\varphi
-\frac{1}{2}m^{2}\varphi^{2}-\eta\varphi^{4}. \label{kappap4}%
\end{equation}
Substituting the replacement (\ref{kappap4}) into the field equation
(\ref{P41}) gives
\begin{equation}
\square\varphi+m^{2}\varphi+4\eta\varphi^{3}=0,
\end{equation}
which is the field equation with $U\left(  \varphi\right)  =\eta\varphi^{4}$.

The duality transformation of the solution is straightforward. Substituting
the solution (\ref{p4sol}) into Eq. (\ref{Ephi}) gives%
\begin{equation}
G=\frac{1}{2}\partial_{\mu}\phi\partial^{\mu}\phi+\frac{m^{2}}{2}\phi
^{2}+\lambda\varphi^{4}=-\frac{m^{4}}{16\lambda}.
\end{equation}
Substituting the transformations (\ref{phi4phitrans}) and (\ref{xtoyphi4})
into the solution (\ref{p4sol}) and using Eq. (\ref{Eeta}) give the solution
of $U\left(  \varphi\right)  =\frac{m^{4}}{16\lambda}\varphi^{4}$:%
\begin{equation}
\varphi=-\frac{2\sqrt{\lambda}}{im}\coth\left(  \alpha\tau+\beta y_{1}+\gamma
y_{2}-\frac{y_{3}}{2}\sqrt{4\alpha^{2}-4\beta^{2}-4\gamma^{2}-2m^{2}}%
-\delta\right)  . \label{varphisol1}%
\end{equation}
Putting $\frac{m^{4}}{16\lambda}=\eta$, we obtain the solution of $U\left(
\varphi\right)  =\eta\varphi^{4}$:%
\begin{align}
\varphi &  =\frac{im}{2\sqrt{\eta}}\coth\left(  \alpha\tau+\beta y_{1}+\gamma
y_{2}-\frac{y_{3}}{2}\sqrt{4\alpha^{2}-4\beta^{2}-4\gamma^{2}-2m^{2}}%
-\delta\right) \nonumber\\
&  =\frac{im}{2\sqrt{\eta}}\tanh\left(  \alpha\tau+\beta y_{1}+\gamma
y_{2}-\frac{y_{3}}{2}\sqrt{4\alpha^{2}-4\beta^{2}-4\gamma^{2}-2m^{2}}%
+\delta^{\prime}\right)
\end{align}
with the constant $\delta^{\prime}=-\delta+i\frac{\pi}{2}$.

\subsection{The $\phi^{1}$-field and the $\phi^{-2}$-field}

The dual field of\ the $\phi^{1}$-field, by the duality relation
(\ref{power}), is the $\phi^{-2}$-field, i.e., the fields
\begin{align}
V\left(  \phi\right)   &  =\lambda\phi,\label{Vphi1}\\
U\left(  \varphi\right)   &  =\eta\varphi^{-2} \label{Uphiwm2}%
\end{align}
are dual to each other. The duality transformations, by Eqs. (\ref{phitans}) ,
(\ref{xtans}), and (\ref{Eeta}), are%
\begin{align}
\phi &  \rightarrow\varphi^{2},\label{Tphiphiw}\\
x^{\mu}  &  \rightarrow2y^{\mu},\label{Txy}\\
\eta &  \rightarrow-G. \label{etamG}%
\end{align}

For the field $\phi$ with the potential (\ref{Vphi1}),%
\begin{equation}
G=\frac{1}{2}\partial_{\mu}\phi\partial^{\mu}\phi+\frac{1}{2}m^{2}\phi
^{2}+\lambda\phi. \label{hphi1}%
\end{equation}

\subsubsection{The $1+3$-dimensional solution}

The field equation of $\phi$, Eq. (\ref{Vphi1}), has a solution%
\begin{equation}
\phi=\exp\left(  \alpha t+\beta x_{1}+\gamma x_{2}+i\sqrt{-m^{2}-\alpha
^{2}+\beta^{2}+\gamma^{2}}x_{3}\right)  -\frac{\lambda}{m^{2}}.
\label{phi4dsol}%
\end{equation}

Substituting the solution (\ref{phi4dsol}) into Eq. (\ref{hphi1}) gives
\begin{equation}
G=-\frac{\lambda^{2}}{2m^{2}}. \label{G1m2}%
\end{equation}

Substituting the duality transformations (\ref{Tphiphiw}) and (\ref{Txy}) into
the solution (\ref{phi4dsol}) and using the duality relation (\ref{etamG}), we
arrive at the solution of $U\left(  \varphi\right)  =\frac{\lambda^{2}}%
{2m^{2}}\varphi^{-2}$:%

\begin{equation}
\varphi=\left[  \exp\left(  2\alpha\tau+2\beta y_{1}+2\gamma y_{2}%
+2i\sqrt{-m^{2}-\alpha^{2}+\beta^{2}+\gamma^{2}}y_{3}\right)  -\frac{\lambda
}{m^{2}}\right]  ^{1/2}. \label{varphi4dsol}%
\end{equation}
This is just the solution of the dual potential $U\left(  \varphi\right)
=\eta\varphi^{-2}$ with $\frac{\lambda^{2}}{2m^{2}}=\eta$:
\begin{equation}
\varphi=\left[  \exp\left(  2\alpha\tau+2\beta y_{1}+2\gamma y_{2}%
+2i\sqrt{-m^{2}-\alpha^{2}+\beta^{2}+\gamma^{2}}y_{3}\right)  -\frac
{\sqrt{2\eta}}{m}\right]  ^{1/2}.
\end{equation}

\subsubsection{The $1+1$-dimensional solution}

The field equation of $\phi$, Eq. (\ref{Vphi1}), has a $1+1$-dimensional
solution,%
\begin{equation}
\phi=e^{\alpha t}\sinh\left(  \sqrt{\alpha^{2}+m^{2}}x\right)  -\frac{\lambda
}{m^{2}}. \label{phi4dsol2}%
\end{equation}
Substituting the solution (\ref{phi4dsol2}) into Eq. (\ref{hphi1}) gives%
\begin{equation}
G=-\frac{\lambda^{2}}{2m^{2}}-\frac{\alpha^{2}+m^{2}}{2}e^{2\alpha t}.
\end{equation}

Substituting the duality transformations (\ref{Tphiphiw}) and (\ref{Txy}) into
the solution (\ref{phi4dsol2}) and using the duality relation (\ref{etamG}),
we arrive at the solution of $U\left(  \varphi\right)  =\left(  \frac
{\lambda^{2}}{2m^{2}}+\frac{\alpha^{2}+m^{2}}{2}e^{2\alpha t}\right)
\varphi^{-2}$:
\begin{equation}
\varphi=\left[  e^{2\alpha\tau}\sinh\left(  2\sqrt{\alpha^{2}+m^{2}}y\right)
-\frac{\lambda}{m^{2}}\right]  ^{1/2}.
\end{equation}

\subsection{The $\phi^{3}$-field and the $\phi^{6}$-field}

The dual field of\ the $\phi^{3}$-field, by the duality relation
(\ref{power}), is the $\phi^{6}$-field, i.e., the fields
\begin{align}
V\left(  \phi\right)   &  =\lambda\phi^{3},\label{Vphi3}\\
U\left(  \varphi\right)   &  =\eta\phi^{6} \label{Uphi6}%
\end{align}
are dual to each other. The duality transformations by Eqs. (\ref{phitans}) ,
(\ref{xtans}), and (\ref{Eeta}) are%
\begin{align}
\phi &  \rightarrow\varphi^{-2},\label{Tphiphiw36}\\
x^{\mu}  &  \rightarrow-2y^{\mu}. \label{Txy36}%
\end{align}
and%
\begin{equation}
\eta\rightarrow-G. \label{etamG36}%
\end{equation}

The field equation of $\phi$, Eq. (\ref{Vphi3}), has a solution%
\begin{equation}
\phi=-\frac{m^{2}}{6\lambda}\left[  3\tanh^{2}\left(  \alpha x_{1}+\beta
x_{2}+\gamma x_{3}+\frac{\sqrt{4\alpha^{2}+4\beta^{2}+4\gamma^{2}+m^{2}}}%
{2}t\right)  -1\right]  . \label{phi34dsol}%
\end{equation}
For the field $\phi$ with the potential (\ref{Vphi3}),%
\begin{equation}
G=\frac{1}{2}\partial_{\mu}\phi\partial^{\mu}\phi+\frac{1}{2}m^{2}\phi
^{2}+\lambda\phi^{3}=\frac{m^{6}}{54\lambda^{2}}. \label{hphi3}%
\end{equation}

Substituting the duality transformations (\ref{Tphiphiw36}) and (\ref{Txy36})
into the solution (\ref{phi34dsol}) and using the duality relation
(\ref{etamG36}), we obtain the solution of $U\left(  \varphi\right)
=-\frac{m^{6}}{54\lambda^{2}}\phi^{6}$:
\begin{equation}
\varphi=\left\{  -\frac{m^{2}}{6\lambda}\left[  3\tanh^{2}\left(  2\alpha
y_{1}+2\beta y_{2}+2\gamma y_{3}+\sqrt{4\alpha^{2}+4\beta^{2}+4\gamma
^{2}+m^{2}}\tau\right)  -1\right]  \right\}  ^{-1/2}. \label{varphi4dsol36}%
\end{equation}
We then arrive at the solution of $U\left(  \varphi\right)  =\eta\phi^{6}$
with $-\frac{m^{6}}{54\lambda^{2}}=\eta$:%
\begin{equation}
\varphi=\left\{  \frac{i\sqrt{6\eta}}{2m}\left[  3\tanh^{2}\left(  2\alpha
y_{1}+2\beta y_{2}+2\gamma y_{3}+\sqrt{4\alpha^{2}+4\beta^{2}+4\gamma
^{2}+m^{2}}\tau\right)  -1\right]  \right\}  ^{-1/2}.
\end{equation}

\subsection{$V\left(  \phi\right)  =\lambda\phi^{3}+\kappa\phi$ and its
duality}

The massless scalar field with the polynomial potential
\begin{equation}
V\left(  \phi\right)  =\lambda\phi^{3}+\kappa\phi\label{Vphi31}%
\end{equation}
has two dual potentials. Taking $a=3$ in the duality relations (\ref{aandA})
and (\ref{bandB}) gives $A=6$ and $B_{n}=4$, i.e., the dual potential is
\begin{equation}
U\left(  \varphi\right)  =\eta\varphi^{6}+\sigma\varphi^{4}. \label{Uvarphi64}%
\end{equation}
Taking $a=1$ in the duality relations (\ref{aandA}) and (\ref{bandB}) gives
$A=-2$ and $B_{n}=4$, i.e., the dual potential is%
\begin{equation}
U\left(  \varphi\right)  =\eta\varphi^{-2}+\sigma\varphi^{4}.
\label{Uvarphi-24}%
\end{equation}
Obviously, the fields with the potentials (\ref{Uvarphi64}) and
(\ref{Uvarphi-24}) are also dual.

The massless field equation of the potential (\ref{Vphi31}) has a solution
\begin{equation}
\phi=i\sqrt{\frac{\kappa}{\lambda}}\left[  \frac{2}{\sqrt{3}}-\sqrt{3}%
\tanh^{2}\left(  \alpha t-\frac{x}{2}\sqrt{4\alpha^{2}-2i\left(
3\kappa\lambda\right)  ^{1/2}}\right)  \right]  . \label{sol31}%
\end{equation}
For the field $\phi$ with the potential (\ref{Vphi31}),%
\begin{equation}
G=\frac{1}{2}\partial_{\mu}\phi\partial^{\mu}\phi+\frac{1}{2}m^{2}\phi
^{2}+\lambda\phi^{3}=\frac{2i\kappa^{3/2}}{3\sqrt{3\lambda}}.
\end{equation}
From the solution of the potential (\ref{Vphi31}), we can obtain the solution
of its dual potentials, Eqs. (\ref{Uvarphi64}) and (\ref{Uvarphi-24}), by
means of the duality relation.

\subsubsection{$U\left(  \varphi\right)  =\eta\varphi^{6}+\sigma\varphi^{4}$}

The duality transformations given by Eqs. (\ref{phitanspol}), (\ref{xtanspol}%
), (\ref{Geta}), and (\ref{kapsig}) are
\begin{align}
\phi &  \rightarrow\varphi^{-2},\nonumber\\
x^{\mu}  &  \rightarrow-2y^{\mu},\nonumber\\
\eta &  \rightarrow-G,\nonumber\\
\sigma &  \rightarrow\kappa.
\end{align}
Substituting the duality relation into the solution (\ref{sol31}) gives the
solution of $U\left(  \varphi\right)  =-\frac{2i\kappa^{3/2}}{3\sqrt{3\lambda
}}\varphi^{6}+\kappa\varphi^{4}$:
\begin{equation}
\varphi=i\sqrt{\frac{\kappa}{\lambda}}\left[  \frac{2}{\sqrt{3}}-\sqrt{3}%
\tanh^{2}\left(  -2\alpha\tau+y\sqrt{4\alpha^{2}-2i\left(  3\kappa
\lambda\right)  ^{1/2}}\right)  \right]  ^{-1/2}. \label{sol641}%
\end{equation}
This is the solution of the dual potential $U\left(  \varphi\right)
=\eta\varphi^{6}+\sigma\varphi^{4}$ when $-\frac{2i\kappa^{2}}{3\sqrt
{3\kappa\lambda}}=\eta$ and $\kappa=\sigma$:
\begin{equation}
\varphi=\left(  \frac{\sigma}{3\eta}\right)  ^{1/2}\left[  1-\frac{3}{2}%
\tanh^{2}\left(  -2\alpha\tau+2y\sqrt{\alpha^{2}+\frac{\sigma^{2}}{3\eta}%
}\right)  \right]  ^{-1/2}.
\end{equation}

\subsubsection{$U\left(  \varphi\right)  =\eta\varphi^{-2}+\sigma\varphi^{4}$}

Similarly, the solution of the potential (\ref{Uvarphi-24}) can be achieved by
the duality transformations (\ref{phitanspol}), (\ref{xtanspol}),
(\ref{Geta}), and (\ref{kapsig}):%
\begin{align}
\phi &  \rightarrow\varphi^{2},\nonumber\\
x^{\mu}  &  \rightarrow2y^{\mu},\nonumber\\
\eta &  \rightarrow-G,\nonumber\\
\sigma &  \rightarrow\lambda.
\end{align}
Substituting into the solution (\ref{sol31}) gives the solution of $U\left(
\varphi\right)  =-\frac{2i\kappa^{2}}{3\sqrt{3\kappa\lambda}}\varphi
^{-2}+\lambda\varphi^{4}$,%
\begin{equation}
\varphi=\left[  2i\sqrt{\frac{\kappa}{3\lambda}}-i\sqrt{\frac{3\kappa}%
{\lambda}}\tanh^{2}\left(  2\alpha\tau-y\sqrt{4\alpha^{2}-2i\left(
3\kappa\lambda\right)  ^{1/2}}\right)  \right]  ^{1/2}. \label{sol-24}%
\end{equation}
This is the solution of the dual potential $U\left(  \varphi\right)
=\eta\varphi^{-2}+\sigma\varphi^{4}$ when $-\frac{2i\kappa^{2}}{3\sqrt
{3\kappa\lambda}}=\eta$ and $\sigma=\lambda$:%
\begin{equation}
\varphi=\left(  \frac{\eta}{2\sigma}\right)  ^{1/6}\left[  3\tanh^{2}\left(
2\alpha\tau-y\sqrt{4\alpha^{2}+6\left(  \frac{\sigma^{2}\eta}{2}\right)
^{1/3}}\right)  -2\right]  ^{1/2}.
\end{equation}

\subsection{$V\left(  \phi\right)  =\lambda\phi^{n}+\kappa_{1}\phi
^{2n-2}+\kappa_{2}$ and its duality}

The massive scalar field with the polynomial potential%
\begin{equation}
V\left(  \phi\right)  =\lambda\phi^{n}+\kappa_{1}\phi^{2n-2}+\kappa_{2},\text{
\ \ }n\geq3, \label{Vphin}%
\end{equation}
has the following dual potentials. By the duality relations (\ref{aandA}) and
(\ref{bandB}), $a=n$ corresponds to $A=\frac{2n}{n-2}$, $B_{1}=-2$ and
$B_{2}=\frac{2n}{n-2}$, i.e., the dual potential is%
\begin{equation}
U\left(  \varphi\right)  =\eta\varphi^{\frac{2n}{n-2}}+\sigma_{1}\varphi
^{-2}+\sigma_{2}\varphi^{\frac{2n}{n-2}}; \label{Uphin1}%
\end{equation}
$a=2n-2$ corresponds to $A=\frac{2n-2}{n-2}$, $B_{1}=1$, and $B_{2}%
=\frac{2n-2}{n-2}$, i.e., the dual potential is%
\begin{equation}
U\left(  \varphi\right)  =\eta\varphi^{\frac{2n-2}{n-2}}+\sigma_{1}%
\varphi+\sigma_{2}\varphi^{\frac{2n-2}{n-2}}. \label{Uphin2}%
\end{equation}
The fields with the potentials (\ref{Uphin1}) and (\ref{Uphin2}) are also dual.

The field equation with the potential (\ref{Vphi31}) has a solution
\cite{polyanin2016handbook}%
\begin{equation}
\phi=\left[  \frac{\sqrt{\lambda^{2}-2\kappa_{1}m^{2}}}{m^{2}}\cos\left(
m\left(  2-n\right)  \left(  t\cosh\alpha+x\sinh\alpha\right)  +\beta\right)
-\frac{\lambda}{m^{2}}\right]  ^{\frac{1}{2-n}}. \label{soln2n-2}%
\end{equation}
For the field $\phi$ with the potential (\ref{Vphin}),%
\begin{equation}
G=\frac{1}{2}\partial_{\mu}\phi\partial^{\mu}\phi+\frac{1}{2}m^{2}\phi
^{2}+\lambda\phi^{3}=\kappa_{2}.
\end{equation}
From the solution of the potential (\ref{soln2n-2}), we can obtain the
solution of its dual potentials (\ref{Uphin1}) and (\ref{Uphin2}) by means of
the duality relation.

\subsubsection{$U\left(  \varphi\right)  =\kappa_{1}\varphi^{-2}$}

The duality transformations given by Eqs. (\ref{phitanspol}), (\ref{xtanspol}%
), (\ref{Geta}), and (\ref{kapsig}) are%
\begin{align}
\phi &  \rightarrow\varphi^{\frac{2}{2-n}},\nonumber\\
x^{\mu}  &  \rightarrow\frac{2}{2-n}y^{\mu},\nonumber\\
\eta &  \rightarrow-G,\nonumber\\
\sigma_{1}  &  \rightarrow\kappa_{1},\nonumber\\
\sigma_{2}  &  \rightarrow\kappa_{2}.
\end{align}
Substituting into the solution (\ref{soln2n-2}) gives the solution of
$U\left(  \varphi\right)  =-\kappa_{2}\varphi^{\frac{2n}{n-2}}+\kappa
_{1}\varphi^{-2}+\kappa_{2}\varphi^{\frac{2n}{n-2}}=\kappa_{1}\varphi^{-2}$:
\begin{equation}
\varphi=\left[  \frac{\sqrt{\lambda^{2}-2\kappa_{1}m^{2}}}{m^{2}}\cos\left(
2m\left(  \tau\cosh\alpha+y\sinh\alpha\right)  +\beta\right)  -\frac{\lambda
}{m^{2}}\right]  ^{1/2}. \label{sol22-n}%
\end{equation}

\subsubsection{$U\left(  \varphi\right)  =\lambda\varphi$}

Similarly, the solution of the potential (\ref{Uphin2}) can be achieved by the
duality transformation given by Eqs. (\ref{phitanspol}), (\ref{xtanspol}),
(\ref{Geta}), and (\ref{kapsig}):
\begin{align}
\phi &  \rightarrow\varphi^{\frac{1}{2-n}},\nonumber\\
x^{\mu}  &  \rightarrow\frac{1}{2-n}y^{\mu},\nonumber\\
\eta &  \rightarrow-G,\nonumber\\
\sigma_{1}  &  \rightarrow\lambda,\nonumber\\
\sigma_{2}  &  \rightarrow\kappa_{2}.
\end{align}

Substituting into the solution (\ref{soln2n-2}) gives the solution of

$U\left(  \varphi\right)  =-\kappa_{2}\varphi^{\frac{2n-2}{n-2}}%
+\lambda\varphi+\kappa_{2}\varphi^{\frac{2n-2}{n-2}}=\lambda\varphi$:%
\begin{equation}
\varphi=\frac{\sqrt{\lambda^{2}-2\kappa_{1}m^{2}}}{m^{2}}\cos\left(  m\left(
\tau\cosh\alpha+y\sinh\alpha\right)  +\beta\right)  -\frac{\lambda}{m^{2}}.
\label{sol12-n}%
\end{equation}

\section{Conclusion \label{conclusion}}

In this paper, we reveal a duality between fields. The dual fields are related
by the duality relation.

It is shown that fields who are dual to each other form a duality family. For
polynomial potentials, the duality family consists of finite number of dual
fields, for nonpolynomial potentials, e.g., the sine-Gordon field, the duality
family consists of infinite number of dual fields. Some fields, e.g., the free
field and the $\phi^{4}$-field, are self-duality.

The existence of the duality family inspires us to classify fields based on
the duality. A duality family is a duality class. In a duality class, the
different of fields is only a duality transformation.

The duality of fields provides a high-efficiency approach for solving field
equations. Once one field equation is solved, all of its dual fields are
solved by the duality relation. That is, in a duality family, we only need to
solve one of its member.

As examples, we solve the $\phi^{-2}$-field from the solution of its dual
field, $\phi^{1}$-field, solve the field equation of the dual field of the
sine-Gordon field from the sine-Gordon field, etc.

In further work, we will consider the quantum theory of dual fields. For
example, the relation of the Feynman rule of dual fields. Especially, in
quantum field theory we will consider the duality in the heat kernel method
\cite{vassilevich2003heat} and in the scattering spectrum method
\cite{graham2009spectral,pang2012relation,li2015heat}. In these methods we can
calculate the one-loop effective action and the vacuum
energy\ \cite{dai2009number,dai2010approach}. We may observe the relation of
the one-loop effective action and the vacuum energy of dual fields. Moreover,
we will consider the duality of spinor fields and vector fields.

\appendix

\section{A solution of the scalar field equation \label{appendix}}

In this appendix we show the field equation%
\begin{equation}
\square\phi+m^{2}\phi+\frac{\partial V\left(  \phi\right)  }{\partial\phi}=0,
\label{motequ}%
\end{equation}
has an implicit solution%
\begin{equation}
\beta_{\mu}x^{\mu}+\int\frac{\sqrt{-\beta^{2}}}{\sqrt{2\left[  \frac{1}%
{2}m^{2}\phi^{2}+V\left(  \phi\right)  -G\right]  }}d\phi\equiv F_{1}\left(
x^{\mu},\phi\right)  =0. \label{gensol}%
\end{equation}

From the solution (\ref{gensol}) we obtain
\begin{equation}
\partial_{\mu}\phi=-\frac{\frac{\partial F_{1}\left(  x^{\mu},\phi\right)
}{\partial x^{\mu}}}{\frac{\partial F_{1}\left(  x^{\mu},\phi\right)
}{\partial\phi}}=-\frac{\beta_{\mu}\sqrt{2\left[  \frac{1}{2}m^{2}\phi
^{2}+V\left(  \phi\right)  -G\right]  }}{\sqrt{-\beta^{2}}}. \label{rdphi1}%
\end{equation}

Furthermore, by Eq. (\ref{rdphi1}), letting%
\begin{equation}
\partial_{\mu}\phi+\frac{\beta_{\mu}\sqrt{2\left[  \frac{1}{2}m^{2}\phi
^{2}+V\left(  \phi\right)  -G\right]  }}{\sqrt{-\beta^{2}}}\equiv F_{2}\left(
x^{\mu},\phi,\partial_{\mu}\phi\right)  =0,
\end{equation}
we arrive at
\begin{align}
\square\phi &  =\partial^{\mu}\partial_{\mu}\phi=\frac{\partial}{\partial
x_{\mu}}\partial_{\mu}\phi\nonumber\\
&  =-\frac{\frac{\partial F_{2}\left(  x^{\mu},\phi,\partial_{\mu}\phi\right)
}{\partial x_{\mu}}}{\frac{\partial F_{2}\left(  x^{\mu},\phi,\partial_{\mu
}\phi\right)  }{\partial_{\mu}\phi}}-\frac{\frac{\partial F_{2}\left(  x^{\mu
},\phi,\partial_{\mu}\phi\right)  }{\partial\phi}}{\frac{\partial F_{2}\left(
x^{\mu},\phi,\partial_{\mu}\phi\right)  }{\partial_{\mu}\phi}}\frac
{\partial\phi}{\partial x_{\mu}}\nonumber\\
&  =-\frac{\partial F_{2}\left(  x^{\mu},\phi,\partial_{\mu}\phi\right)
}{\partial\phi}\partial^{\mu}\phi\nonumber\\
&  =-\left(  m^{2}\phi+\frac{\partial V\left(  \phi\right)  }{\partial\phi
}\right)  .
\end{align}
This is the field equation (\ref{motequ}).

\acknowledgments

We are very indebted to Dr G. Zeitrauman for his encouragement. This work is supported in part by NSF of China under Grant
No. 11575125 and No. 11675119.

%(正文结束)――――――――――――――――――――――――――――――――――――――――――――――――――

%%%%%%%%%%%%%%%%%%%参考文献%%%%%%%%%%%%%%%%%%%%%%%%%%%

%%%%%%%%%%%%%%%%%%%bibtex形式的参考文献%%%%%%%%%%%%%%%
%\bibliographystyle{JHEP} %参考文献的风格(.bst)
%\bibliography{refs} %参考文献文件(.bib)
%\nocite{*} %若去掉注释，没有被引用的文献也被列出

%%%%%%%%%%%%%%%%%%%bbl形式的参考文献%%%%%%%%%%%%%%%%%%

%%%%%%%%%%%%%%%%%%%%%%%%%%%%%%%%%%%%%%%%%%%%%%%%%%%%%%

\end{CJK*}
\end{document}